\documentclass[12pt,a4paper]{article}

\usepackage{amsmath,amssymb,slashed,cancel}
\usepackage{cite,tabularx}
\usepackage{subcaption}
\usepackage{graphicx,color}
\usepackage[normalem]{ulem}
\usepackage{mathtools}
\usepackage{enumerate}
\usepackage{ascmac}
\usepackage{enumitem}
\usepackage{float}
\usepackage{placeins}
\allowdisplaybreaks

\usepackage[height=8.85in,width=6.4in]{geometry}
\renewcommand{\baselinestretch}{1.1}
\usepackage{comment}
\newcommand\package[2][\relax]{\texttt{#2\ifx#1\relax\relax\relax\else\,\linebreak[0]#1\fi}}

\numberwithin{equation}{section} 
 
\def\beq#1\eeq{\begin{align}#1\end{align}}

\definecolor{BlueViolet}{rgb}{0.2, 0.00, 0.7}
\definecolor{Blue}{rgb}{0.15, 0.00, 0.9}
\usepackage[colorlinks=true,linkcolor=Blue,citecolor=Blue,urlcolor=BlueViolet]{hyperref}

\begin{document}
\begin{titlepage}
\setcounter{page}{0} 

\begin{center}

\vskip .55in

\begingroup
\centering
\large\bf Interplay between decoupling and non-decoupling effective field theories in electroweak phase transitions
\endgroup

\vskip .4in

{
Katsuya Hashino$^{\rm (a)}$ and Daiki Ueda$^{\rm (b)}$
}

\vskip 0.4in

\begingroup\small
\begin{minipage}[t]{0.9\textwidth}
\centering\renewcommand{\arraystretch}{0.9}
\begin{tabular}{c@{\,}l}
$^{\rm(a)}$
& National Institute of Technology, Fukushima College, Nagao 30\\
&Taira-Kamiarakawa, Iwaki, Fukushima 970–8034, Japan
\\[2mm]
$^{\rm(b)}$
& Physics Department, Technion \text{--} Israel Institute of Technology
\\
&Technion city, Haifa 3200003, Israel\\
\end{tabular}
\end{minipage}
\endgroup

\end{center}

\vskip .4in

\begin{abstract}\noindent
Gravitational-wave (GW) observations offer a promising probe of new physics associated with a strong first-order electroweak phase transition. 
Null results from direct searches and constraints from precision measurements, together with the comparatively weak constraints on Higgs self-interactions, motivate scenarios in which non-decoupling effects in the Higgs potential coexist with additional decoupling new physics.
Previous Fisher matrix studies within the Standard Model effective field theory (SMEFT) have explored the sensitivity of future GW observations to new physics, focusing on transitions driven by the dimension-six operator $(H^\dagger H)^3$.
However, the validity of the truncated EFT expansion can become questionable in the parameter regions relevant to such transitions.
We instead consider transitions driven by non-decoupling effects described by the nearly aligned Higgs effective field theory (naHEFT).
We investigate whether GW observations can probe small corrections to the Higgs effective potential induced by decoupling effects in interactions other than Higgs self-interactions.
As a representative benchmark for additional decoupling effects, we consider the dimension-six SMEFT operator $(H^\dagger H)(\overline{q}_3 t_R\widetilde{H})$ with Wilson coefficient $C_{uH}$.
Although we focus on this benchmark, the same framework can be applied to other dimension-six SMEFT operators. 
We perform two-parameter Fisher matrix analyses for DECIGO and BBO to quantify their expected sensitivity to $C_{uH}$ and the naHEFT mass scale $\Lambda$.
For the benchmark configurations considered, the non-decoupling dynamics generates a detectable GW signal.
We find that these small corrections can produce potentially measurable changes in the GW spectrum, even after accounting for the correlation between $C_{uH}$ and $\Lambda$.

\end{abstract}
\end{titlepage}

\setcounter{page}{1}
\renewcommand{\thefootnote}{\#\arabic{footnote}}
\setcounter{footnote}{0}

\begingroup
\renewcommand{\baselinestretch}{1} 
\setlength{\parskip}{2pt}          
\hrule
\tableofcontents
\vskip .2in
\hrule
\vskip .4in
\endgroup

\section{Introduction}\label{sec:intro}

The Higgs boson was discovered at the CERN Large Hadron Collider (LHC)~\cite{ATLAS:2012yve,CMS:2012qbp}, and measurements of its properties have so far been consistent with the predictions of the Standard Model (SM).
Nevertheless, the shape of the Higgs potential has not yet been fully determined experimentally, and the nature of the electroweak phase transition therefore remains one of the major open questions in particle physics.
In particular, a strong first-order electroweak phase transition (SFO-EWPT) can provide the necessary departure from thermal equilibrium for electroweak baryogenesis~\cite{Gavela:1993ts,Konstandin:2003dx,Kuzmin:1985mm}, generate a stochastic gravitational-wave background~\cite{Grojean:2006bp,Kakizaki:2015wua}, and, under suitable conditions, lead to the formation of primordial black holes~\cite{Liu:2021svg,Hashino:2021qoq,Hashino:2025fse}.
In the SM, however, the electroweak transition is a smooth crossover for the measured Higgs-boson mass of approximately $125\,\mathrm{GeV}$~\cite{Kajantie:1995kf,Kajantie:1996mn,Kajantie:1996qd,Csikor:1998eu,DOnofrio:2015gop}.
A SFO-EWPT therefore requires new physics (NP) that modifies the Higgs potential.
At the same time, the absence of direct evidence for new particles at the LHC motivates the investigation of scenarios in which the relevant NP scale lies above the electroweak symmetry-breaking scale.

These observations motivate an effective field theory (EFT) description, in which the effects of heavy NP are encoded in higher-dimensional operators obtained by integrating out the new degrees of freedom.
It is widely expected that the SM constitutes the leading renormalizable part of an EFT valid below a cutoff scale associated with NP.
The Standard Model Effective Field Theory (SMEFT)~\cite{Grzadkowski:2010es,Jenkins:2013zja,Jenkins:2013wua,Alonso:2013hga} extends the SM Lagrangian by including higher-dimensional operators constructed from the SM fields and consistent with the SM gauge symmetries.
When the NP scale is well separated from the electroweak scale and its low-energy effects exhibit decoupling, the SMEFT expansion can be systematically organized in inverse powers of the NP scale and truncated at a fixed mass dimension.
The phenomenological effects of dimension-six SMEFT operators have been extensively studied in a wide range of observables, including collider processes, and stringent constraints have been placed on their Wilson coefficients~\cite{deBlas:2025xhe,ATLAS:2026fyh,CMS:2026zfh}.
The absence of significant deviations from the SM predictions is therefore compatible with NP scenarios characterized by a sufficiently high scale or sufficiently small couplings.

Within this framework, the effects of higher-dimensional SMEFT operators on the Higgs potential and the realization of a SFO-EWPT have been extensively studied~\cite{Ellis:2018mja,Grojean:2004xa,Zhang:1992fs,Bodeker:2004ws,Binetruy:2012ze,Delaunay:2007wb,Huber:2013kj,Konstandin:2014zta,Damgaard:2015con,Harman:2015gif,Balazs:2016yvi,deVries:2017ncy,Cai:2017tmh,Chala:2018ari,Dorsch:2018pat,DeVries:2018aul,Chala:2019rfk,Ellis:2019flb,Zhou:2019uzq,Kanemura:2020yyr,Phong:2020ybr,Wang:2020zlf,Wang:2020jrd,Kanemura:2021fvp,Lewicki:2021pgr,Hashino:2021qoq,Kanemura:2022txx,Anisha:2022hgv,Huang:2015izx,Cao:2017oez,Huang:2016odd,Croon:2020cgk,Hashino:2022ghd,Banerjee:2024qiu,Gazi:2024boc,Camargo-Molina:2024sde,Bernardo:2025vkz,Chala:2025aiz,Zhu:2025pht,Chala:2025gif,Braathen:2026jfg,Abu-Ajamieh:2026ndt,Liu:2026ask,Liu:2025ipj,Jahedi:2025yjz,Hashino:2025nku}.
Of particular relevance to the present study, Refs.~\cite{Hashino:2022ghd,Hashino:2025nku} investigated the effects of various dimension-six and selected dimension-eight SMEFT operators on the Higgs potential, the dynamics of the SFO-EWPT, and the resulting GW spectrum, together with their detectability in future GW observations.
The analysis in Ref.~\cite{Hashino:2025nku} provided a comprehensive treatment of dimension-six operators, with particular emphasis on uncertainties associated with the choice of renormalization scale.
These studies showed that, provided the \((H^\dagger H)^3\) operator induces a SFO-EWPT accompanied by a potentially observable GW signal, the resulting GW spectrum can retain sensitivity to a variety of other dimension-six operators even after renormalization-scale uncertainties are taken into account.

The dimension-six SMEFT operator \((H^\dagger H)^3\) provides a systematic and largely model-independent parametrization of leading NP effects on the Higgs potential and has been extensively studied in the context of a SFO-EWPT.
However, realizing a SFO-EWPT through this operator can push the SMEFT expansion toward the boundary of its regime of validity, thereby calling into question the reliability of a truncation at dimension six.
In Ref.~\cite{Damgaard:2015con}, the validity of this description was examined by comparing the electroweak phase transition in a singlet extension of the SM with that predicted by the corresponding EFT obtained after integrating out the singlet scalar.
It was found that the EFT reproduces the behavior of the full model only within a restricted region of the singlet-model parameter space.
A more systematic analysis was performed in Ref.~\cite{Postma:2020toi}, where various UV extensions were matched onto the SMEFT using the covariant derivative expansion.
The analysis showed that a SMEFT truncated at dimension six generally does not provide an accurate description of a SFO-EWPT, with a limited exception for singlet extensions admitting tree-level matching; even in this case, dimension-eight operators can be required for quantitatively reliable predictions.
These observations motivate considering frameworks that allow for non-decoupling NP effects in the Higgs sector, while the effects relevant to other sectors may remain decoupling.

Motivated by the possibility that the SFO-EWPT is driven by non-decoupling NP effects, the nearly aligned Higgs effective field theory (naHEFT)~\cite{Kanemura:2021fvp,Kanemura:2022txx} has been proposed as a model-independent parametrization of a broad class of such effects.
In the naHEFT framework, deviations in the Higgs interactions are assumed to arise predominantly from loop effects of additional particles, whose non-decoupling contributions to the Higgs potential are parametrized in a Coleman--Weinberg-like form.
The naHEFT has been applied to various phenomena associated with a SFO-EWPT, including the resulting stochastic GW background and the formation of primordial black holes~\cite{Kanemura:2022txx,Hashino:2022tcs,Florentino:2024kkf,Hashino:2025fse}.
These developments in the EFT description of non-decoupling NP, together with the extensive studies of decoupling NP effects within the SMEFT, naturally raise the following question: can future GW observations probe additional decoupling NP effects when the SFO-EWPT itself is driven by non-decoupling dynamics?

In this work, we address this question by investigating how the interplay between decoupling and non-decoupling EFTs affects the SFO-EWPT and the resulting GW spectrum.
We describe the non-decoupling dynamics responsible for the SFO-EWPT within the naHEFT framework, while parametrizing additional decoupling NP effects in terms of dimension-six SMEFT operators.
In contrast to Refs.~\cite{Hashino:2022ghd,Hashino:2025nku}, where the SFO-EWPT is induced by the dimension-six SMEFT operator \((H^\dagger H)^3\), we consider a scenario in which the phase transition is driven primarily by the non-decoupling effects encoded in the naHEFT.
To illustrate the impact of additional decoupling NP, we focus on the dimension-six operator $
(H^\dagger H)(\overline{q}_i u_j\widetilde{H})$ as a representative benchmark.
The effects of this and other dimension-six SMEFT operators on the effective Higgs potential have been studied in Refs.~\cite{Hashino:2022ghd,Hashino:2025nku}.
Using these results, we can straightforwardly extend our analysis to other dimension-six SMEFT operators.
Ref.~\cite{Kanemura:2022txx} identified regions of the naHEFT parameter space in which a SFO-EWPT produces GW signals within the sensitivity reach of DECIGO.
Focusing on these regions, we perform a Fisher matrix analysis to quantify the sensitivity of future GW observations to the additional decoupling contribution represented by $
(H^\dagger H)(\overline{q}_i u_j\widetilde{H})$.
We show that future GW observations can retain sensitivity to decoupling NP effects even when the SFO-EWPT and the associated GW signal are primarily generated by non-decoupling dynamics.
Our results therefore demonstrate that the qualitative conclusions of Refs.~\cite{Hashino:2022ghd,Hashino:2025nku} can persist beyond scenarios in which the phase transition itself is induced by SMEFT operators.

This paper is organized as follows.
In Section~\ref{sec:form}, we introduce the EFT framework employed in this work, which combines non-decoupling effects described by the naHEFT with decoupling effects parametrized by a dimension-six SMEFT operator, and present the corresponding Higgs effective potential.
In Section~\ref{sec:num}, we numerically study the effects of these contributions on the SFO-EWPT and the resulting GW spectrum.
We then perform a Fisher matrix analysis to quantify the sensitivity of future GW observations to the decoupling contribution in the presence of non-decoupling dynamics, and examine how the projected sensitivity depends on the assumed bubble-wall velocity.
Our main numerical results are presented in Figs.~\ref{fig:cont}--\ref{fig:k4r05}.
Finally, we summarize our results and discuss their implications in Section~\ref{sec:summary}.

\section{Higgs effective potential}\label{sec:form}
To investigate the interplay between decoupling and non-decoupling new physics, we construct an EFT description of the Higgs effective potential that incorporates both types of effects.
Specifically, we employ the naHEFT~\cite{Kanemura:2022txx} to describe the non-decoupling contributions, while parametrizing the decoupling contributions in terms of dimension-six SMEFT operators.
In Section~\ref{sec:naHEFT}, we introduce this combined EFT framework and derive the corresponding tree-level parameter-matching conditions.
In Section~\ref{sec:matching_loop}, we extend these matching conditions to include the one-loop contributions.
Section~\ref{sec:CW} presents the zero-temperature one-loop Coleman--Weinberg potential, while Section~\ref{sec:thermal} introduces the finite-temperature effective potential and summarizes the procedure used to evaluate the SFO-EWPT within the naHEFT framework.

\subsection{EFT for decoupling and non-decoupling physics}
\label{sec:naHEFT}
Throughout this paper, we consider an EFT described by the Lagrangian
\begin{align}
\mathcal{L}
=
\mathcal{L}_{\rm SM}
+
\mathcal{L}_{\rm NP}\,,
\label{eq:full_L}
\end{align}
where \(\mathcal{L}_{\rm SM}\) denotes the SM Lagrangian and \(\mathcal{L}_{\rm NP}\) encodes the effects of additional heavy particles that have been integrated out.
The possibility of an SFO-EWPT is phenomenologically motivated by its potential role in electroweak baryogenesis and by the possible production of a stochastic GW background and primordial black holes.
Non-decoupling effects in the Higgs potential provide a well-motivated mechanism for realizing such a transition.
At the same time, the absence of compelling evidence for NP in collider experiments is consistent with scenarios in which additional NP contributions to interactions other than Higgs self-interactions are decoupling in nature.
These considerations motivate studying the coexistence of non-decoupling effects in the Higgs potential and additional decoupling NP.
We therefore adopt the phenomenological decomposition
\begin{align}
\mathcal{L}_{\rm NP}
=
\mathcal{L}_{\rm non}
+
\mathcal{L}_{\rm dec}\,,
\label{eq:NP}
\end{align}
where \(\mathcal{L}_{\rm non}\) describes the dominant non-decoupling contributions to the Higgs dynamics that drive the SFO-EWPT, while \(\mathcal{L}_{\rm dec}\) parametrizes additional decoupling contributions in terms of SMEFT operators.
Here, non-decoupling refers to effects that do not vanish in the heavy-mass limit, as can occur when the masses of the new particles arise partly from electroweak symmetry breaking.
Such effects generally cannot be represented reliably by a finite truncation of the SMEFT expansion.
By contrast, decoupling effects are systematically suppressed by inverse powers of the heavy NP scale and can be organized as a truncated SMEFT expansion when a sufficient separation of scales is present.

For the decoupling contributions, we follow Refs.~\cite{Hashino:2022ghd,Hashino:2025nku} and work at dimension six in the SMEFT expansion:
\begin{align}
\mathcal{L}_{\rm dec}
=
\sum_i C_i\,\mathcal{O}_i^{(6)}\,,
\label{eq:dec}
\end{align}
where \(\mathcal{O}_i^{(6)}\) are dimension-six operators constructed from the SM fields and invariant under the SM gauge symmetries, and the Wilson coefficients \(C_i\) have mass dimension \(-2\).
Refs.~\cite{Hashino:2022ghd,Hashino:2025nku} investigated the effects of various such operators on the phase-transition dynamics and the resulting GW spectrum, assuming that the SFO-EWPT is induced by the operator \((H^\dagger H)^3\).
In the present work, by contrast, the phase transition is driven primarily by the non-decoupling Higgs dynamics encoded in \(\mathcal{L}_{\rm non}\), and we investigate whether the resulting GW spectrum retains sensitivity to an additional decoupling contribution.
As a representative benchmark, we consider the CP-even third-generation operator
\begin{align}
\mathcal{L}_{\rm dec}
\supset
C_{uH}\,(H^\dagger H)
(\overline{q}_3 t_R\widetilde{H})
+\mathrm{h.c.}\,,
\qquad
C_{uH}\in\mathbb{R}\,.
\label{eq:bench}
\end{align}
This operator modifies the field-dependent top-quark mass and consequently affects both the zero-temperature and finite-temperature effective potentials through top-quark loops.
A comprehensive analysis of other decoupling operators is beyond the scope of this work, although their effects can be incorporated into the present framework following the methods developed in Refs.~\cite{Hashino:2022ghd,Hashino:2025nku}.

To study non-decoupling NP, we adopt the naHEFT framework introduced in Ref.~\cite{Kanemura:2022txx}.
In this framework, the effects of NP are parameterized by
\begin{align}
\mathcal{L}_{\rm non}= -\frac{\xi}{4}\kappa_0 \left[\mathcal{M}^2(H)\right]^2\ln \frac{\mathcal{M}^2(H)}{\overline{\mu}^2}\,,\label{eq:LP}
\end{align}
where $|H|^2=(v+h)^2/2$, with $v=1/(2^{1/4}\sqrt{G_F})\simeq 246$ GeV and $h$ denoting the Higgs field. The parameters $\kappa_0$ and $\overline{\mu}^2$ are real and have mass dimensions zero and two, respectively, while $\xi=1/(4\pi)^2$. The function $\mathcal{M}^2(H)$ is an arbitrary function of $|H|^2$ with mass dimension two.
Equation~\eqref{eq:LP} incorporates the one-loop effects of NP.
Following Ref.~\cite{Kanemura:2022txx}, we parametrize $\mathcal{M}^2(H)$ as follows for numerical convenience:
\begin{align}
    \mathcal{M}^2(H)=M^2+\kappa_p |H|^2\,,
\end{align}
where $M^2$ and $\kappa_p$ are real parameters.
For later convenience, we define
\begin{align}
    \Lambda^2&\equiv M^2+\frac{\kappa_p}{2}v^2\,,\qquad r\equiv \frac{\frac{\kappa_p}{2}v^2}{\Lambda^2}=1-\frac{M^2}{\Lambda^2}\,.\label{eq:Lam_r}
\end{align}
The above expressions can be rewritten as
\begin{align}
    M^2&= \left(1-r\right)\Lambda^2\,,\qquad\kappa_p=\frac{2 \Lambda^2}{v^2}r\,.
\end{align}
The parameter $\Lambda$ corresponds to the mass scale of the heavy new particle integrated out of the theory, while $r$ characterizes the size of the non-decoupling effects.
In the decoupling limit, $r\ll 1$, $\mathcal{L}_{\rm non}$ admits a valid EFT expansion.
For example, expanding the non-renormalizable terms in $\mathcal{L}_{\rm non}$ up to dimension ten yields
\begin{align}
    \mathcal{L}_{\rm non}\supset -\frac{\xi}{4}\kappa_0\bigg[\frac{\kappa_p^3}{3M^2}|H|^6 - \frac{\kappa_p^4}{12M^4} |H|^8
    +
    \frac{\kappa_p^5 }{30 M^6}|H|^{10}+\mathcal{O}\left(|H|^{12}\right)
    \bigg]\,.
\end{align}
In the decoupling limit, the resulting EFT can be matched onto the Higgs self-interaction sector of the SMEFT.

The non-decoupling contributions encoded in Eq.~\eqref{eq:LP} arise at the one-loop level and modify the parameters of the Higgs potential through loop corrections.
The one-loop effects, including those described by the naHEFT, are discussed mainly in Section~\ref{sec:matching_loop}.
The Higgs-field-dependent terms in Eq.~\eqref{eq:full_L} take the form
\begin{align}
    \mathcal{L}\supset \mathcal{L}_{\rm Higgs}&\equiv(D_{\mu}H)^{\dagger}(D^{\mu}H) +m^2 H^{\dagger}H -\frac{\lambda}{2} (H^{\dagger}H)^2-\frac{\xi}{4}\kappa_0 [\mathcal{M}^2(H)]^2\ln \frac{\mathcal{M}^2(H)}{\overline{\mu}^2}
    \,,\label{eq:SMEFT_Higgs}
\end{align}
where the covariant derivative is defined by $D_{\mu} = \partial_{\mu} + i g_s T^A G^A_{\mu} + i g T^I W^I_{\mu} + i g' Y B_{\mu}$,  
with $\{g_s,\, g,\, g'\}$ being the gauge couplings and $\{G_{\mu}^A,\, W_{\mu}^I,\, B_{\mu}\}$ the gauge fields of the $SU(3)_c$, $SU(2)_L$, and $U(1)_Y$, respectively. 
The Higgs-sector Lagrangian takes the form
\begin{align}
    \mathcal{L}_{\rm Higgs}\supset \frac{1}{2}(\partial_{\mu}h)^2-V_h\,,
\end{align}
where the potential $V_h$ is given by
\begin{align}
    V_h=V+\Delta {V}_h\,.\label{eq:Vh}
\end{align}
Here, $V$ denotes the $h$-independent part of the potential and is given by
\begin{align}
    V=-\frac{m^2}{2}\phi^2+\frac{\lambda}{8}\phi^4+\frac{\xi}{4}\kappa_0 [\mathcal{M}^2(\hat{H})]^2\ln \frac{\mathcal{M}^2(\hat{H})}{\overline{\mu}^2}\,,\label{eq:Vtree}
\end{align}
where $|\hat{H}|^2= \phi^2/2$ with $\phi=v$.
The parameters $m^2$ and $\lambda$ can be determined using the following relations:
\begin{align}
    \partial_h V_h|_{h=0}=0\,,\qquad
    \partial_h^2 V_h|_{h=0}=M_h^2 \,,\label{eq:ddV}
\end{align}
where $M_h$ denotes the Higgs boson mass.
Treating the naHEFT contribution as a one-loop effect, we obtain the following tree-level relations from Eq.~\eqref{eq:ddV}:
$
(m^2)_{\rm tree}=\frac{1}{2}M_h^2,\,
(\lambda)_{\rm tree}=\frac{M_h^2}{v^2},
$
where the right-hand sides are expressed in terms of the physical quantities $M_h$ and $v$.
In the next section, we extend these tree-level relations to include both the naHEFT contribution and the one-loop SM corrections.

\subsection{Matching including one-loop effects}
\label{sec:matching_loop}
The one-loop SM corrections and NP effects induce shifts in the parameters entering the effective Higgs potential.
The corresponding $\overline{\rm MS}$ parameters are given by
\begin{align}
    m_{\phi}^2 &=M^2_h +{\rm Re}\, \Pi_h(M^2_h)\,,\label{eq:mphi}
    \\
    m^2_W &= M^2_W +{\rm Re}\, \Pi_W (M^2_W)\,,\label{eq:mW}
    \\
    m^2_t &= M^2_t \left(1+2 {\rm Re}\, \Sigma_t(M^2_t)\right)\,,\label{eq:mr}
    \\
    g^2 &=g_{0}^2 +\delta g^2\,,\label{eq:g2}
\end{align}
where $m_{\phi}$, $m_W$, and $m_t$ denote the $\overline{\rm MS}$ masses of the Higgs boson, the $W$ boson, and the top quark, respectively, and $g$ denotes the $SU(2)_L$ gauge coupling in the $\overline{\rm MS}$ scheme.
The quantities $M_h$, $M_W$, and $M_t$ denote the corresponding pole masses.
The tree-level gauge coupling $g_0$, determined from the measured value of $G_F$, is defined by
$
g_0^2 \equiv 4\sqrt{2}\,G_F M_W^2.
$
Here, the subscript $0$ denotes quantities evaluated at tree level.
Furthermore, $\Pi_h$, $\Pi_W$, and $\Sigma_t$ denote the one-loop self-energies of the Higgs boson, the $W$ boson, and the top quark, respectively, while $\delta g^2$ denotes the one-loop correction to the $SU(2)_L$ gauge coupling.
The SM and NP contributions to the one-loop quantities entering Eqs.~\eqref{eq:mphi}--\eqref{eq:g2} are given by
\begin{align}
    \Pi_h(M^2_h)&=\Pi_{h,\,{\rm SM}}(M^2_h)+\Pi_{h,\,{\rm NP}}(M^2_h)\,,\label{eq:Pih}
    \\
    \Pi_W(M^2_W)&=\Pi_{W,\,{\rm SM}}(M^2_W)+\Pi_{W,\,{\rm NP}}(M^2_W)\,,
    \\
    \Sigma_t(M^2_t)&=\Sigma_{\rm SM}(M_t^2)+\Sigma_{\rm NP}(M_t^2)\,,
    \\
    \frac{\delta g^2}{g_{0}^2}&=\left(\frac{\delta g^2}{g_{0}^2}\right)_{\rm SM}+\left(\frac{\delta g^2}{g_{0}^2}\right)_{\rm NP}\,,\label{eq:delg2}
\end{align}
where the superscripts SM and NP denote the SM and NP contributions, respectively.
The SM contributions are given in Refs.~\cite{Kajantie:1995dw,Hashino:2025nku}.
Using Eqs.~\eqref{eq:Pih}--\eqref{eq:delg2}, the $\overline{\rm MS}$ parameters including the one-loop corrections can be expressed as
\begin{align}
    m^2&=\frac{1}{2}\left(M^2_h+{\rm Re}\, \Pi_h\left(M^2_h\right)\right)\,,
    \\
    \lambda&=\frac{M^2_h}{v^2}
    +
    \frac{1}{4}g_0^2 \frac{M^2_h}{M^2_W}
    \left[
    \frac{\delta g^2}{g_0^2}+
    \frac{{\rm Re}\, \Pi_h\left(M^2_h\right)}{M^2_h}
    -
    \frac{{\rm Re}\, \Pi_W\left(M^2_W\right)}{M^2_W}
    \right]\,,
    \\
    Y^2_t&=\frac{1}{2}g^2_0 \frac{M^2_t}{M^2_W}
    \left[1+\frac{\delta g^2}{g_0^2}-
    \frac{{\rm Re}\, \Pi_W\left(M^2_W\right)}{M^2_W}
    +
    2{\rm Re}\,\Sigma_t\left(M^2_t\right)
    \right]\,.
\end{align}
For clarity, the matching relations displayed above include only the SM and naHEFT contributions. 
Although not displayed explicitly, the SMEFT contributions to the matching relations relevant to the \(C_{uH}\) benchmark are included in our numerical analysis, following Ref.~\cite{Hashino:2025nku}.
In what follows, we present the non-decoupling contributions to the quantities in Eqs.~\eqref{eq:Pih}--\eqref{eq:delg2}.
Using Eq.~\eqref{eq:SMEFT_Higgs}, we obtain the non-decoupling contributions to the quantities in Eqs.~\eqref{eq:Pih}--\eqref{eq:delg2} as follows:
\begin{align}
\Pi_{h,{\rm NP}} \left(M_h^2\right)&\supset
-\frac{1}{2} \kappa_0  \kappa_p \xi \left(\kappa_p v^2-M^2-2M^2
\ln \frac{M^2+\frac{\kappa_p v^2}{2}}{\overline{\mu}^2}
\right)\,,
\\
\Pi_{W,{\rm NP}} \left(M_W^2\right)&\supset 
\frac{1}{4}\kappa_0 \kappa_p \xi\frac{M_W^2}{M_h^2} \left(2M^2 +\kappa_p v^2\right)
\left(1+2 \ln \frac{M^2+\frac{\kappa_p v^2}{2}}{\overline{\mu}^2}\right)\,,
\\
\Sigma_{\rm NP}\left(M_t^2\right)&\supset \frac{1}{8}\kappa_0 \kappa_p \xi
\frac{1}{M_h^2} \left(2 M^2+\kappa_p v^2\right)
\left(1+2 \ln \frac{M^2+\frac{\kappa_p v^2}{2}}{\overline{\mu}^2}\right)\,.
\end{align}
Using Eqs.~\eqref{eq:mphi}--\eqref{eq:g2}, the $\overline{\rm MS}$ parameters are determined at one-loop order in terms of physical input parameters.
In our analysis, the one-loop matching is performed at the renormalization scale $\overline{\mu}=M_Z$.
In contrast to Ref.~\cite{Hashino:2025nku}, we do not include RGE effects in our main analysis.
In Appendix~\ref{sec:rge}, however, we discuss the impact of RGE effects and the associated renormalization-scale dependence on our results.

\subsection{CW potential}
\label{sec:CW}
At one-loop order, the Coleman--Weinberg (CW) potential~\cite{Coleman:1973jx} also contributes to the effective Higgs potential.
The one-loop contribution to the CW potential is given by
\begin{align}
    V_{\rm CW}=\sum_{i}\frac{n_i}{64\pi^2}M_i^4\left(\phi\right) \left(
    \ln \left(\frac{M^2_i \left(\phi\right)}{\overline{\mu}^2}\right)-c_i
    \right),\label{eq:CW}
\end{align}
with $\overline{\mu}$ denoting the $\overline{\rm MS}$ renormalization scale.
In the following analysis, except in Appendix~\ref{sec:rge}, all quantities on the right-hand side are evaluated at $\overline{\mu}=M_Z$.
The sum runs over the Higgs boson $h$, the Nambu--Goldstone (NG) modes, the $W$ and $Z$ bosons, and the top quark $t$.
The factor $n_i$ denotes the number of degrees of freedom of each particle species, with
$n_h=1$, $n_{1+}=n_{1-}=1$, $n_{2+}=n_{2-}=2$, $n_W=6$, $n_Z=3$, and $n_t=-12$.
Furthermore, $c_i=5/6$ for gauge bosons and $c_i=3/2$ for the other fields.
Including the NP contributions, we compute the field-dependent masses entering Eq.~\eqref{eq:CW}.
Explicit expressions for the field-dependent masses are provided in Appendix~\ref{sec:FDM}.
The decoupling effects are discussed in detail in Ref.~\cite{Hashino:2025nku}.

\subsection{Effective potential with thermal effects}
\label{sec:thermal}
Finally, we incorporate finite-temperature effects into the Higgs potential derived above.
The finite-temperature contribution to the effective Higgs potential is given by
\begin{align}
    V_T=\frac{T^4}{2\pi^2}
    \left\{
    \sum_i n_i I_{\rm B} \left(\left(M_i(\phi)/T\right)^2 \right)
    + n_t I_{\rm F} \left(\left(M_t(\phi)/T\right)^2 \right)
    \right\}\,,\label{eq:VT}
\end{align}
where the numbers of degrees of freedom $n_i$ and the field-dependent masses $M_i^2(\phi)$ are the same as those appearing in Eq.~\eqref{eq:CW}.
In addition to the SM bosonic contributions, the naHEFT contribution enters through the first term in the curly brackets in Eq.~\eqref{eq:VT}, with $n_{\rm naHEFT}=\kappa_0$ and $M^2_{\rm naHEFT}(\phi)=\mathcal{M}^2(\hat{H})$, assuming that the heavy new particles are bosonic.
The thermal functions~\cite{Dolan:1973qd} are given by
\begin{align}
    I_{\rm B}(a_i^2)&=\int_0^{\infty} dx x^2 \ln \left[1-{\rm exp}\left(-\sqrt{x^2 +a_i^2}\right)\right]\,,
    \\
    I_{\rm F}(a_i^2)&=\int_0^{\infty}dx x^2 \ln \left[
    1+{\rm exp}\left(-\sqrt{x^2 +a_i^2}\right)
    \right]\,,
\end{align}
with $a_i\equiv M_i(\phi)/T$.

Using Eqs.~\eqref{eq:Vtree}, \eqref{eq:CW}, and \eqref{eq:VT}, we obtain the effective potential,
\begin{align}
    V_{\rm full}(\phi,T)= V +V_{\rm CW} + V_T\,.\label{eq:Vfull}
\end{align}
In this paper, to avoid infrared divergences at high temperatures, we employ the Parwani scheme~\cite{Parwani:1991gq}, following the prescription of Ref.~\cite{Hashino:2022tcs}\footnote{An alternative thermal resummation scheme is the Arnold–Espinosa prescription. 
The choice of prescription can affect the predicted high-temperature behavior.
For example, in the two-Higgs-doublet-model parameter regions studied in Ref.~\cite{Bittar:2025lcr}, the Arnold–Espinosa prescription yields symmetry non-restoration, whereas the Parwani prescription yields symmetry restoration.
}.
In this scheme, the field-dependent squared masses of the scalar bosons and the longitudinal gauge-boson modes entering the effective potential in Eq.~\eqref{eq:Vfull} are replaced by their thermally resummed counterparts:
\begin{align}
   M^2_{i}\left(\phi\right)\to M^2_{i,\,{\rm res.}}\left(\phi,T\right)\,.\label{eq:daisy}
\end{align}
In Appendix~\ref{sec:resum}, we present the thermally resummed masses $M^2_{i,,{\rm res.}}(\phi,T)$.
Following Ref.~\cite{Kanemura:2022txx}, the naHEFT contribution is included as follows:
\begin{align}
    M^2_{\rm naHEFT}(\phi)\to M^2_{\rm naHEFT}(\phi)+ \frac{\kappa_p}{6}\,T^2 \,\Theta \left(\kappa_0\right)\,,\label{eq:naHEFT_res} 
\end{align}
where
\begin{align}
    \Theta \left(\kappa_0\right)
    =
    \begin{cases}
    1 & \kappa_0>0
    \\
    0 & \kappa_0\leq 0
    \end{cases}\,.
\end{align}
In the following analysis, we evaluate the EWPT and the generated GWs using Eq.~\eqref{eq:Vfull} together with Eqs.~\eqref{eq:daisy} and \eqref{eq:naHEFT_res}.

\section{Numerical analysis}
\label{sec:num}
We present numerical results following the procedure described in Section~\ref{sec:form}.
As a benchmark scenario, we consider the non-decoupling effects described by the naHEFT in Eq.~\eqref{eq:LP}, together with the decoupling effects induced by a single dimension-six SMEFT operator in Eq.~\eqref{eq:bench}.
A comprehensive analysis of the decoupling effects induced by dimension-six SMEFT operators can be found in Ref.~\cite{Hashino:2025nku}.
Reference~\cite{Hashino:2022tcs} studied the impact of the non-decoupling effects in Eq.~\eqref{eq:LP} on the GW spectrum generated by the SFO-EWPT and showed that the resulting GW signals can be detectable by DECIGO.
Figure~4 of Ref.~\cite{Hashino:2022tcs} identifies the regions of the $(\Lambda,\,\kappa_0,\,r)$ parameter space that yield detectable GW signals.
Focusing on these regions, we investigate the interplay between decoupling and non-decoupling NP effects.

\begin{figure}[H]
  \centering
  \includegraphics[width=7.6cm]{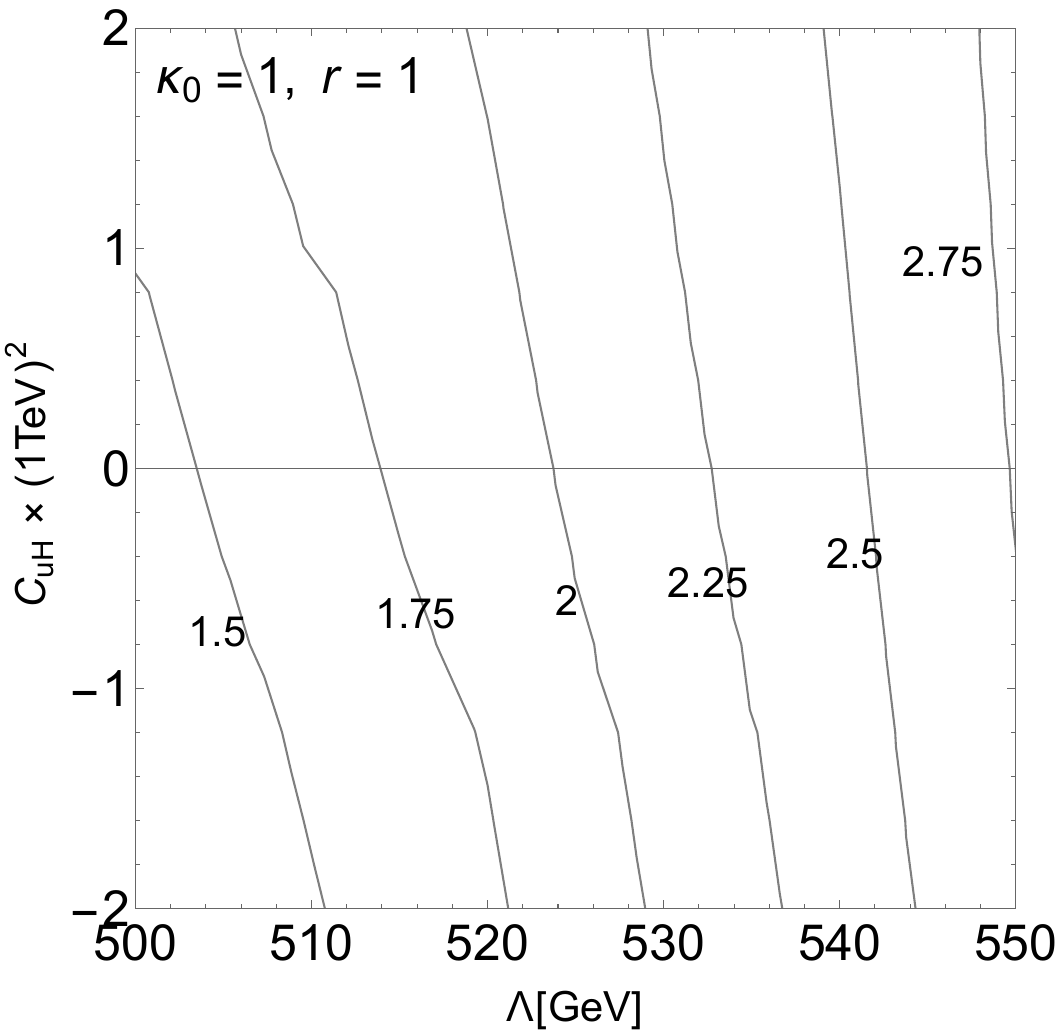}
  \includegraphics[width=7.6cm]{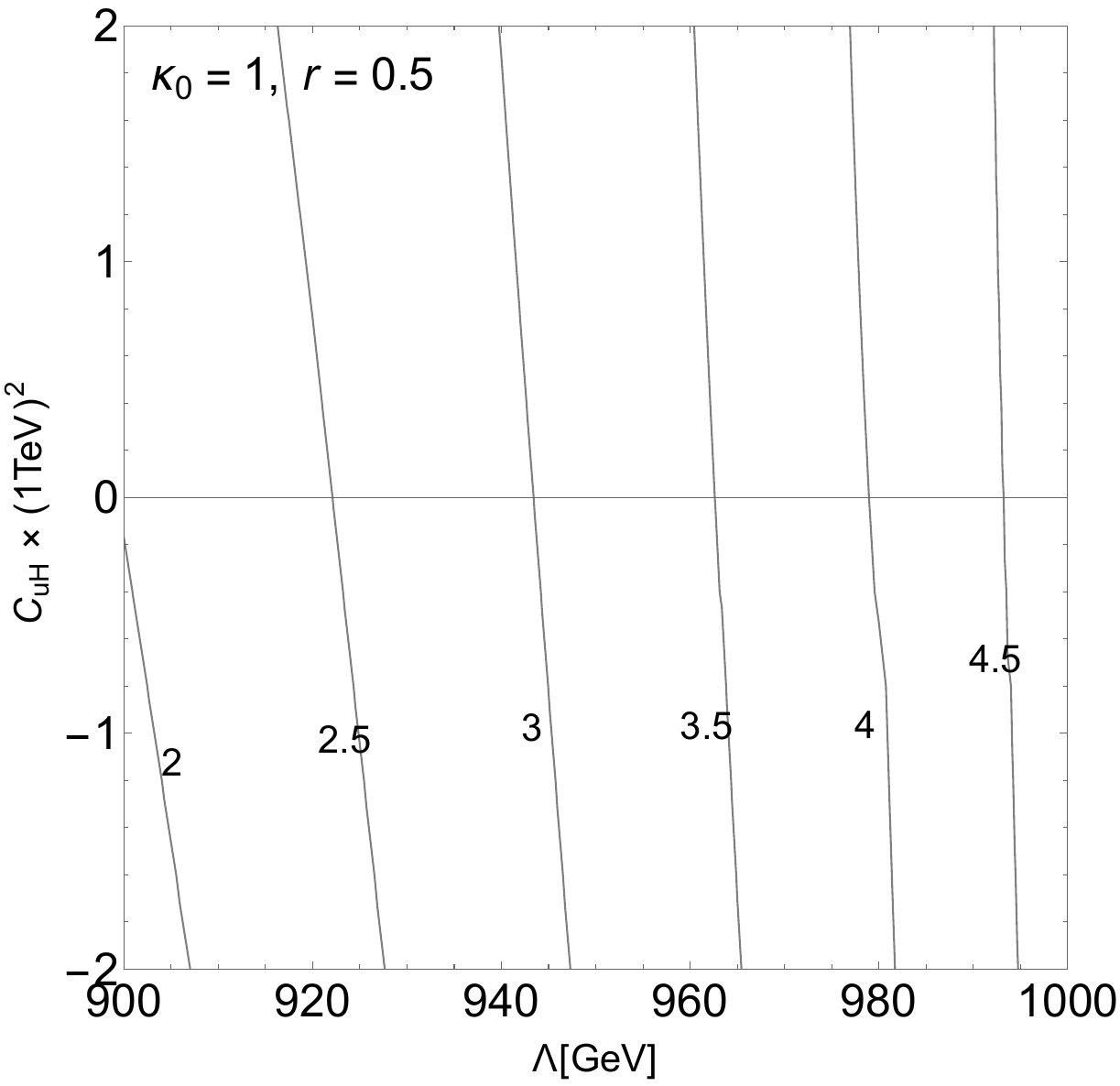}\\
  \includegraphics[width=7.6cm]{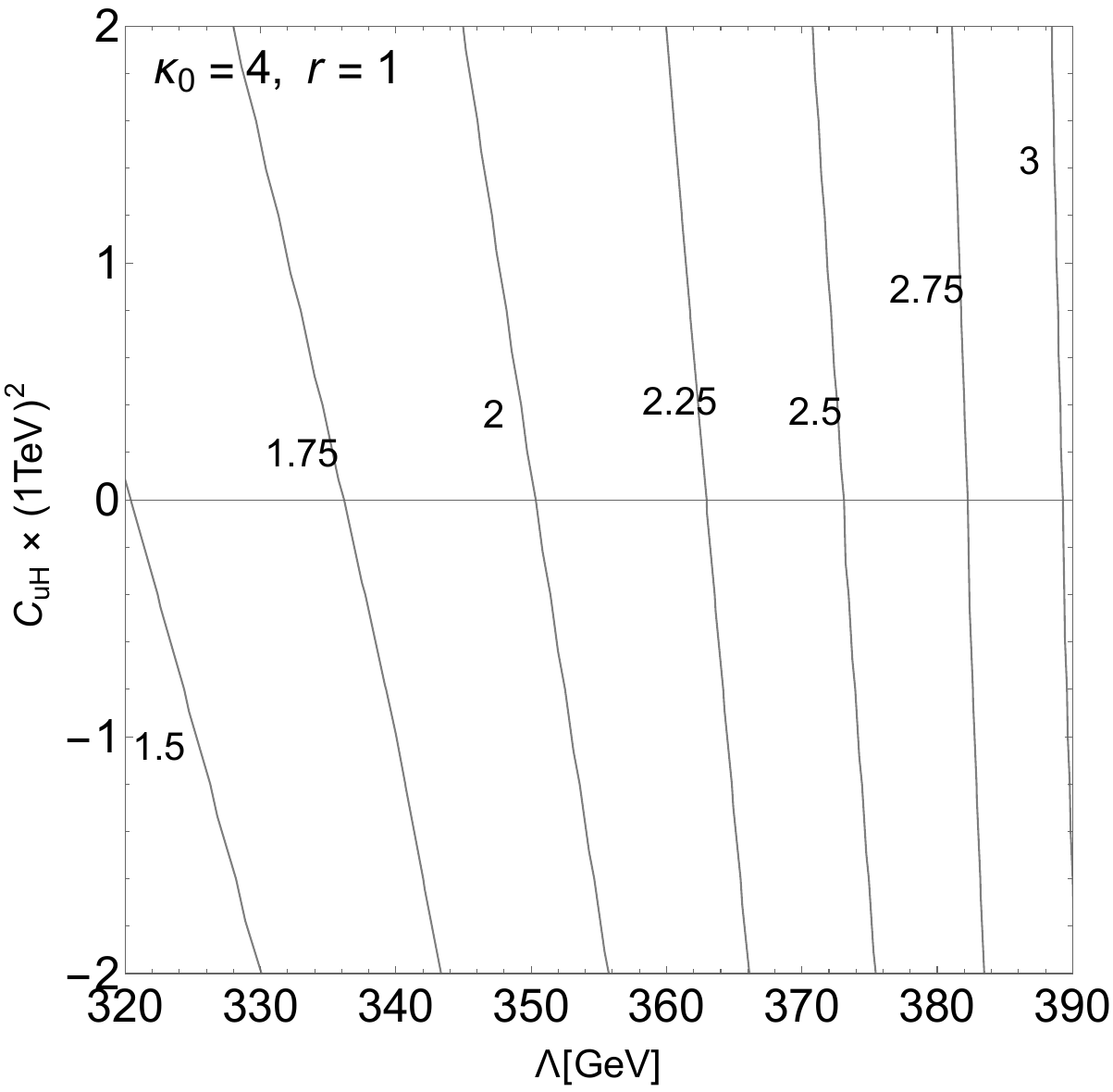}
  \includegraphics[width=7.6cm]{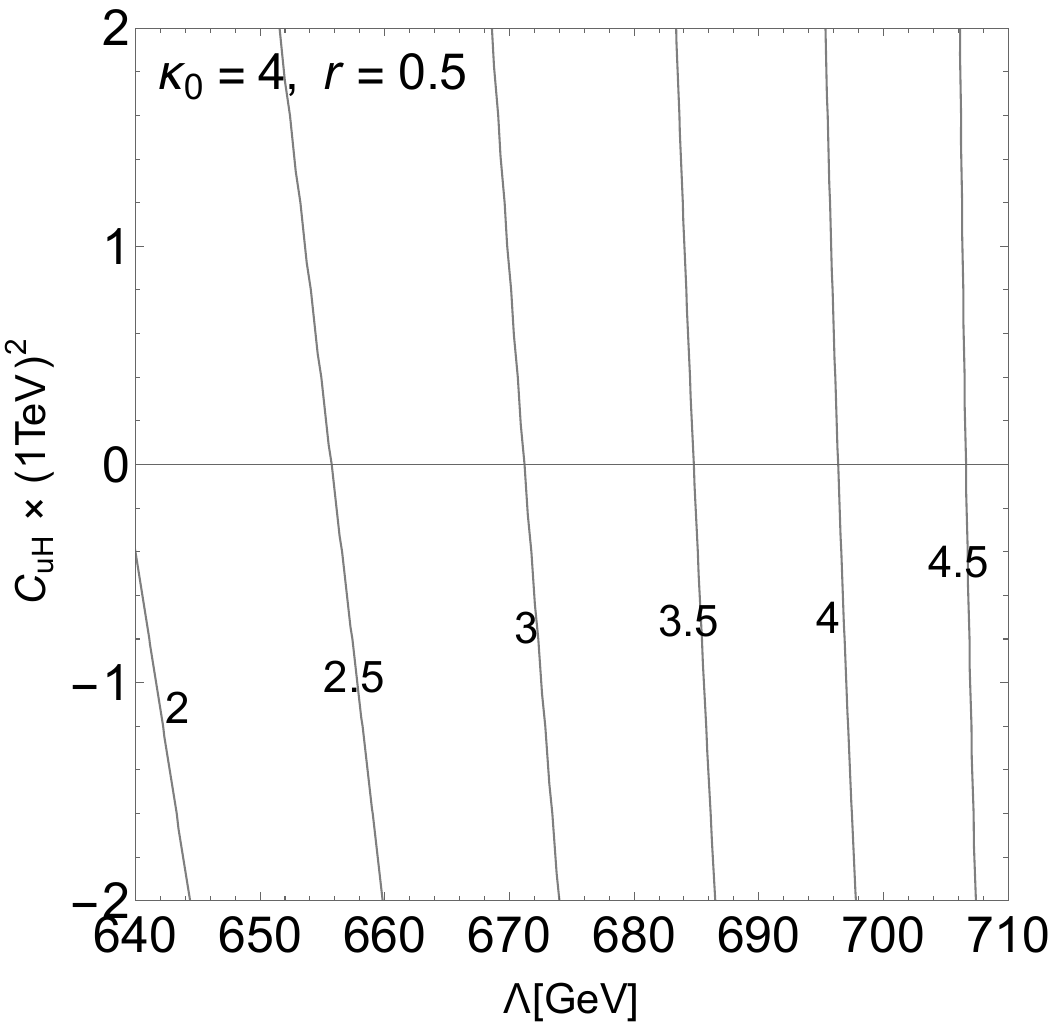}
  \vspace{-0.1cm}
  \caption{Contour plots of $v_c/T_c$ in the
  $(\Lambda,\,C_{uH}\times(1~{\rm TeV})^2)$ plane.
  Here, $C_{uH}$ is the Wilson coefficient defined in
  Eq.~\eqref{eq:bench}, while $\Lambda$ denotes the mass scale
  characterizing the non-decoupling new physics, as defined in
  Eq.~\eqref{eq:Lam_r}.
  The benchmark points are $(\kappa_0,r)=(1,1)$ (upper left),
  $(1,0.5)$ (upper right), $(4,1)$ (lower left), and $(4,0.5)$
  (lower right). These benchmark points are chosen from regions
  where the resulting GW signals are expected to be detectable
  by DECIGO, following Ref.~\cite{Hashino:2022tcs}.}
  \label{fig:vctc}
\end{figure}

We first evaluate the ratio $v_c/T_c$, which characterizes the strength of the SFO-EWPT.
A SFO-EWPT is typically taken to require $v_c/T_c \geq 1$.
The critical temperature $T_c$ and the Higgs VEV at the critical temperature, $v_c$, are determined by the following two conditions:
\begin{align}
    &\partial_{\phi} V_{\rm full}\left(\phi,T_c\right)|_{\phi=v_c}=0\,,
    \\
    &V_{\rm full}\left(v_c,T_c\right)=V_{\rm full}\left(0,T_c\right)\,,
\end{align}
where $V_{\rm full}(\phi,T)$ is given in Eq.~\eqref{eq:Vfull}, with the thermally resummed masses.
In Fig.~\ref{fig:vctc}, we present numerical contour plots of $v_c/T_c$ as functions of $\Lambda$ in Eq.~\eqref{eq:Lam_r} and $C_{uH}\times{\rm (1~TeV)}^2$ for different values of $(\kappa_0,\,r)$.
Figure~\ref{fig:vctc} shows that $v_c/T_c$ is only weakly sensitive to the benchmark decoupling effect parametrized by $C_{uH}$ over the range relevant to current collider constraints~\cite{ATLAS:2026fyh}.
By contrast, the non-decoupling effect plays the dominant role in driving the SFO-EWPT, and therefore primarily controls the value of $v_c/T_c$.
Since the phase-transition parameters are closely related to the strength of the phase transition---for example, $\alpha$, which characterizes the latent heat released during the first-order phase transition, is correlated with $v_c/T_c$~\cite{Espinosa:2010hh}---this result suggests that observables associated with the SFO-EWPT, such as the GW spectrum, are expected to be significantly more sensitive to the non-decoupling effect than to the benchmark decoupling effect.
Although $C_{uH}$ has only a small effect on $v_c/T_c$ over the range considered, its impact on the GW spectrum must be assessed through the nucleation temperature and the other phase-transition parameters.
We therefore examine this sensitivity using the Fisher matrix analysis described below.

The GW spectrum from a first-order phase transition is characterized by the nucleation temperature \(T_n\), the normalized latent heat released by the first-order phase
transition \(\alpha\), the inverse of the duration of the first-order
phase transition \(\beta/H\), and the bubble-wall velocity \(v_b\).
We numerically evaluate \(T_n\), \(\alpha\), and \(\beta/H\) and calculate the resulting GW spectrum following the procedures described in Ref.~\cite{Hashino:2025nku}, while treating \(v_b\) as an input parameter. 
We refer the reader to that reference for the definitions of these parameters and the details of the calculation~\cite{Grojean:2006bp}.

We next present the results of a Fisher matrix analysis to quantify how precisely future GW observations can constrain the NP parameters, including the Wilson coefficient $C_{uH}$ and the non-decoupling parameters, assuming that the SFO-EWPT is driven by the naHEFT.
Once a GW signal is generated by the non-decoupling effects, the decoupling effects can modify its spectrum and may therefore also be probed through GW observations~\cite{Hashino:2022ghd,Hashino:2025nku}.
The Fisher matrix analysis provides a quantitative framework for assessing the interplay and correlations between the decoupling and non-decoupling effects.
We follow the statistical framework adopted in Refs.~\cite{Hashino:2022ghd,Hashino:2025nku}.
To estimate the confidence intervals for the NP parameters from GW observations, we consider the logarithm of the likelihood function, approximated as follows~\cite{Seto:2005qy,Hashino:2018wee,Hashino:2022ghd}:
%
\begin{align}
  \delta \chi^2(\{p\},\{\hat{p}\})\simeq {\cal F}_{ab}(p_a-\hat{p}_a)(p_b-\hat{p}_b)\,,\label{eq:chi}
\end{align}
where $\{p_a\}$ denotes the set of NP parameters, including $C_{uH}$ and $\Lambda$, $\{\hat{p}_a\}$ denotes the corresponding fiducial values, and ${\cal F}_{ab}$ denotes the Fisher information matrix~\cite{Seto:2005qy},
\begin{align}
{\mathcal F}_{ab}
&=
2T_{\rm obs}
\int_0^\infty df
~
\frac{\partial_{p_a} S_h(f,\left\{ \hat{p} \right\}) \partial_{p_b} S_h(f,\left\{ \hat{p} \right\})}
{\left[ S_{\rm eff}(f) + S_h(f,\left\{ \hat{p} \right\}) \right]^2}\,.
\label{eq:FabSeff}
\end{align}
%
Here, $T_{\rm obs}$ denotes the observation time, and $S_h$ is the power spectrum,
%
\begin{align}
S_h(f)
&= 
\frac{3H_0^2}{2\pi^2}
\frac{1}{f^3}
\Omega_{\rm GW}(f)\,,\label{eq:pow}
\end{align}
%
with the GW spectrum $\Omega_{\rm GW}$.
The effective sensitivities $S_{\rm eff}$ for LISA, DECIGO, and BBO are taken from Refs.~\cite{Yagi:2011wg,Klein:2015hvg,Hashino:2022ghd,Hashino:2025nku}.

\begin{figure}[t]
\begin{center}
\includegraphics[width=10.cm]{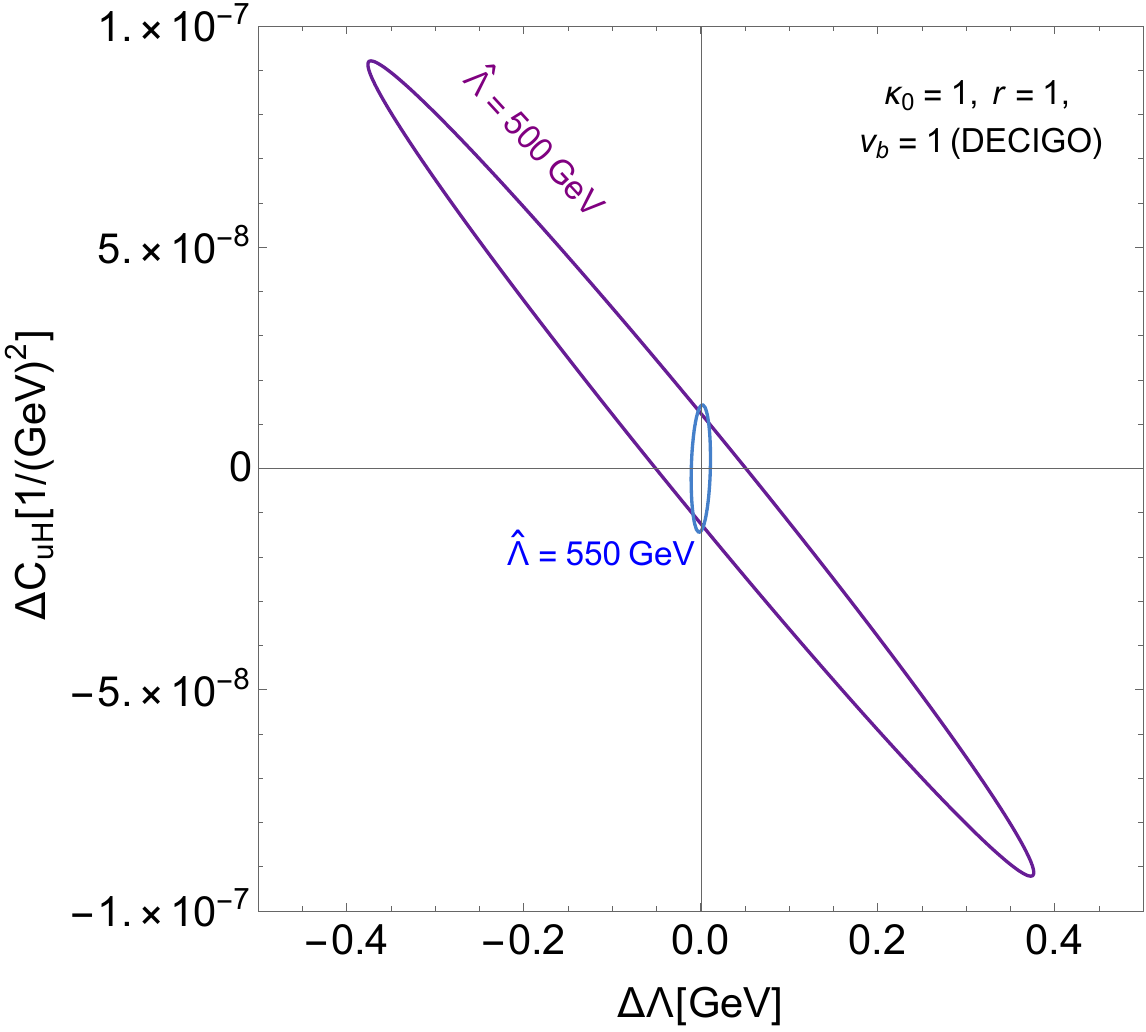}
\end{center}
\vspace{-0.4cm}
\caption{Expected 95\% joint confidence contours for DECIGO, assuming an observation time of one year.
The purple and blue contours correspond to the fiducial values $\hat{\Lambda}=500~{\rm GeV}$ and $550~{\rm GeV}$, respectively, with $\hat{C}_{uH}=0$ in both cases.
The remaining parameters are fixed to $(\kappa_0,\,r)=(1,\,1)$ and $v_b=1$.
Deviations from the fiducial values are defined by $\Delta\Lambda\equiv\Lambda-\hat{\Lambda}$ and $\Delta C_{uH}\equiv C_{uH}-\hat{C}_{uH}$.
}
\label{fig:cont}
\end{figure}

Using Eqs.~\eqref{eq:chi} and \eqref{eq:FabSeff}, we evaluate the expected 95\% confidence intervals for DECIGO and BBO, assuming an observation time of one year ($T_{\rm obs}=1$ year).
The expected joint 95\% confidence regions are defined by $\Delta\chi^2 \leq 5.99$, corresponding to two fitted parameters.
In our setup, the Lagrangian contains four NP parameters: three parameters characterizing the non-decoupling effects, $\Lambda$, $\kappa_0$, and $r$, and one parameter characterizing the decoupling effect, $C_{uH}$.
In addition, the predicted GW spectrum depends on the bubble-wall velocity $v_b$.
Following Ref.~\cite{Hashino:2025nku}, we consider the benchmark values $v_b=0.2$, $0.5$, and $1$ to assess how our results depend on the bubble-wall velocity.
For each choice of $\kappa_0$, $r$, and $v_b$, we vary the fiducial value of $\Lambda$, which largely controls the GW signal, and analyze the confidence intervals in the two-dimensional parameter space $\{p_a\}=\{\Lambda,\,C_{uH}\}$.
The fiducial value of the Wilson coefficient $C_{uH}$ is fixed to zero in order to isolate the sensitivity of GW observations to the decoupling contribution associated with $C_{uH}$.
Figure~\ref{fig:cont} shows the resulting 95\% confidence regions for DECIGO in the $(\Delta\Lambda,\,\Delta C_{uH})$ plane as a representative example, where $\Delta\Lambda\equiv\Lambda-\hat{\Lambda}$ and $\Delta C_{uH} \equiv C_{uH}-\hat{C}_{uH}$.
The purple and blue ellipses correspond to the fiducial points $\{\hat{\Lambda},\,\hat{C}_{uH}\}=\{500~{\rm GeV},\,0~{\rm GeV}^{-2}\}$ and $\{550~{\rm GeV},\,0~{\rm GeV}^{-2}\}$, respectively, with $\kappa_0=1$, $r=1$, and $v_b=1$.
Starting from the fiducial point $(\hat{\Lambda},\,\hat{C}_{uH})=(500~{\rm GeV},\,0)$, increasing $C_{uH}$ while keeping $\Lambda$ fixed changes the predicted GW spectrum and increases $\Delta\chi^2$.
The negative correlation indicates that this increase in $\Delta\chi^2$ can be reduced by appropriately decreasing $\Lambda$.
Allowing $\Lambda$ to vary therefore makes the effect of $C_{uH}$ harder to distinguish from that of $\Lambda$, weakening the expected constraint on $C_{uH}$ compared with the case in which $\Lambda$ is fixed.
Nevertheless, the finite extent of the ellipse shows that, even after $\Lambda$ is adjusted to minimize $\Delta\chi^2$ for each value of $C_{uH}$, only a finite range of $C_{uH}$ remains within the 95\% joint confidence region.
Thus, within the local Fisher approximation and with the remaining parameters fixed, GW observations can constrain $C_{uH}$ even when $\Lambda$ is fitted simultaneously.
The projection of each ellipse onto the $\Delta C_{uH}$ axis gives the range of $ C_{uH}$ contained within the joint confidence region after allowing $\Lambda$ to vary.
The smaller projected extent for $\hat{\Lambda}=550~{\rm GeV}$ therefore corresponds to a tighter expected constraint on $C_{uH}$ at this fiducial point.
These results demonstrate that, within the adopted setup, GW observations can retain sensitivity to the additional decoupling contribution even after accounting for its correlation with the non-decoupling parameter $\Lambda$.
In what follows, we extend this analysis by varying $\kappa_0$, $r$, $v_b$, and the fiducial value of $\Lambda$ to investigate the sensitivity of GW observations to the decoupling effect associated with $C_{uH}$.
%

\begin{figure}[H]
\begin{center}
\includegraphics[width=16.cm]{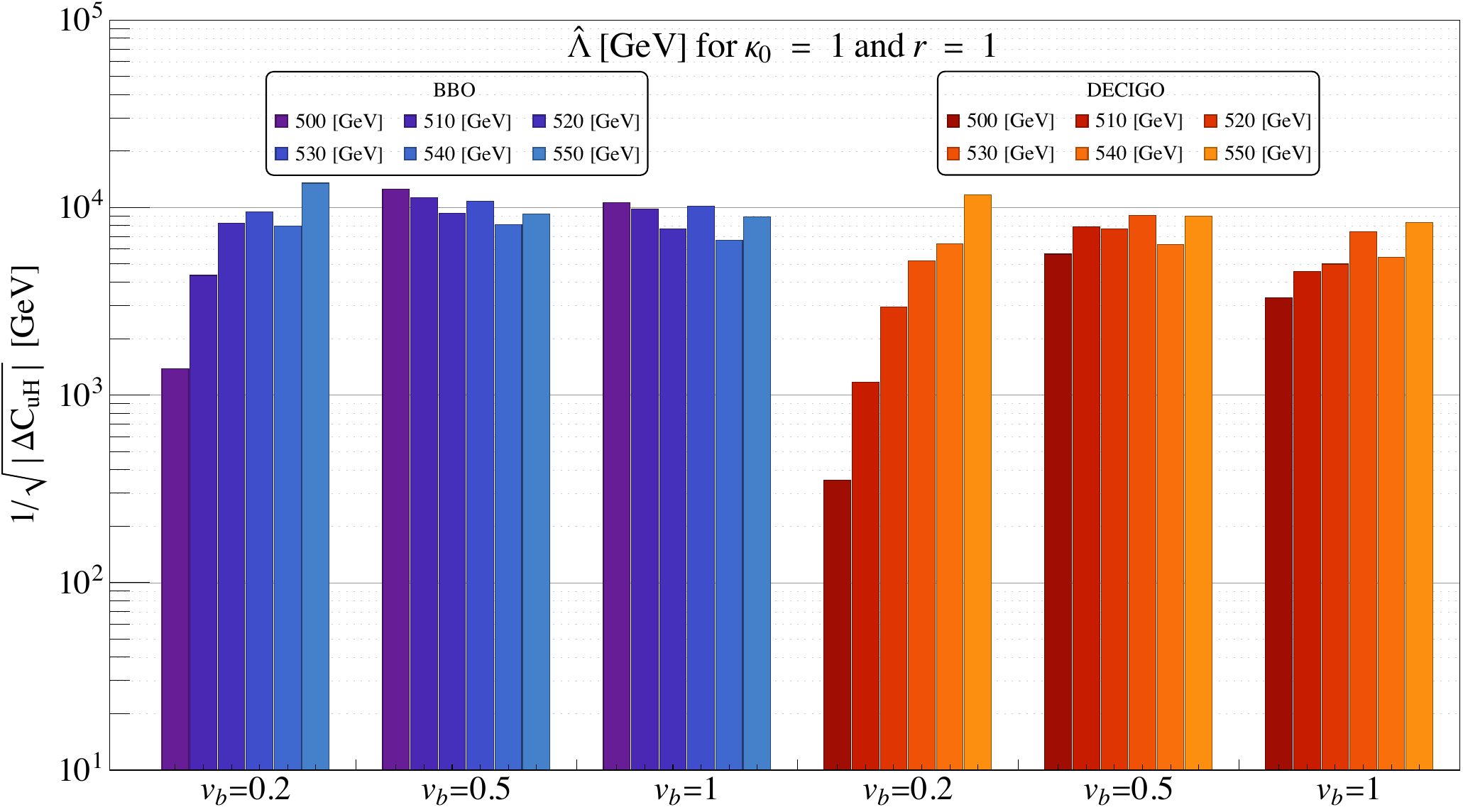}
\end{center}
\vspace{-0.4cm}
\caption{Projected sensitivity to the effective scale $|\Delta C_{uH}|^{-1/2}$ for DECIGO (reddish bands) and BBO (bluish bands), assuming an observation time of one year.
The effective-scale reach is obtained by taking the inverse square root of the upper bound on $|\Delta C_{uH}|$ determined by projecting each expected 95\% joint confidence region onto the $\Delta C_{uH}$ axis, with $\Lambda$ allowed to vary, as illustrated in Fig.~\ref{fig:cont}.
The parameters $(\kappa_0,\,r)$ are fixed to $(1,\,1)$, and the benchmark bubble-wall velocities are $v_b=0.2$, $0.5$, and $1$.
For each experiment, different shades correspond to the fiducial values $\hat{\Lambda}=500$, $510$, $520$, $530$, $540$, and $550~{\rm GeV}$.
Here, $\Delta C_{uH}\equiv C_{uH}-\hat{C}_{uH}$, with $\hat{C}_{uH}=0$ in all cases.
}
\label{fig:k1r1}
\end{figure}

\begin{figure}[H]
\begin{center}
\includegraphics[width=16.cm]{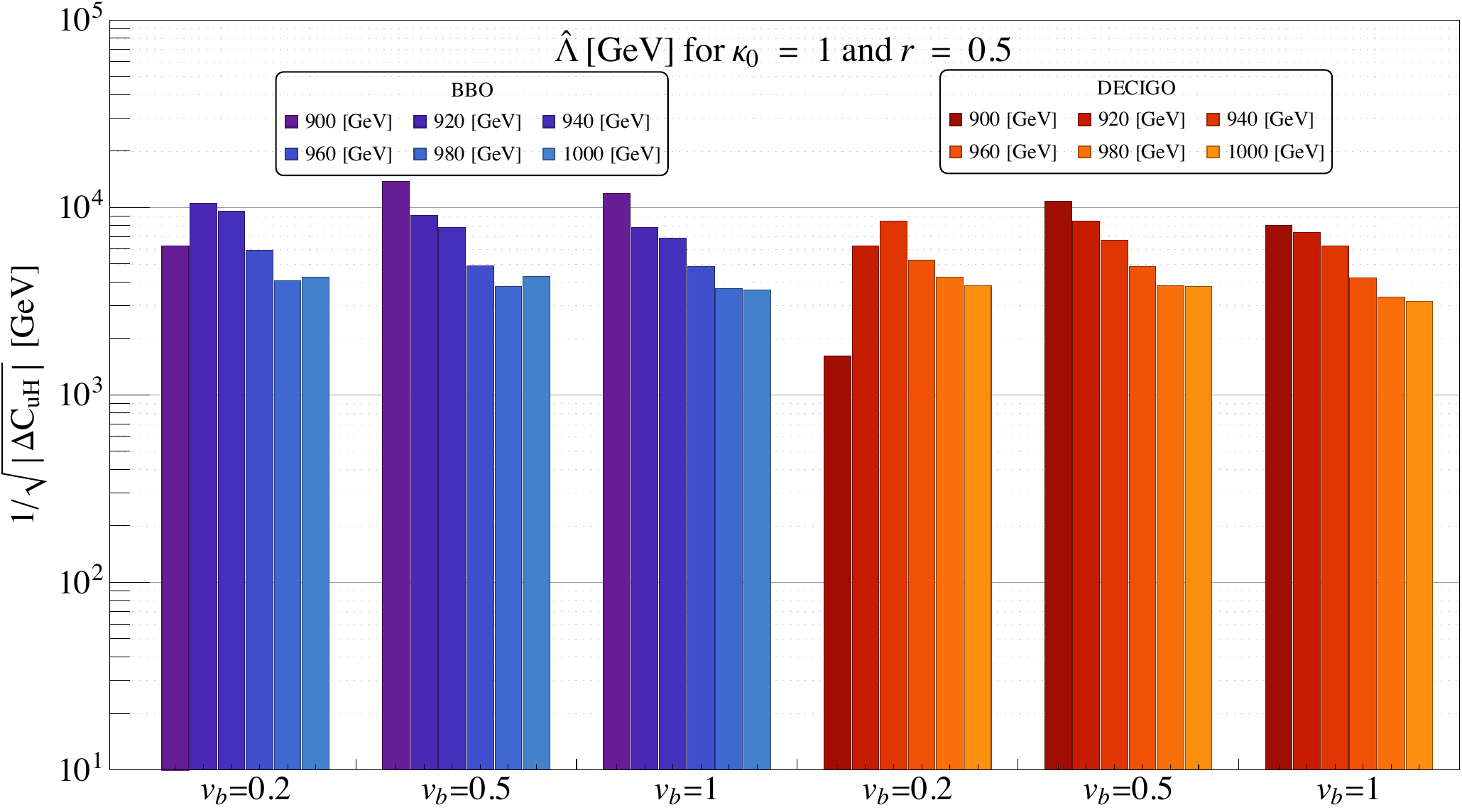}
\end{center}
\vspace{-0.4cm}
\caption{The same plots as Fig.~\ref{fig:k1r1} but for $(\kappa_0\,,\,r)=(1,\,0.5)$. 
}
\label{fig:k1r05}
\end{figure}

\begin{figure}[H]
\begin{center}
\includegraphics[width=16.cm]{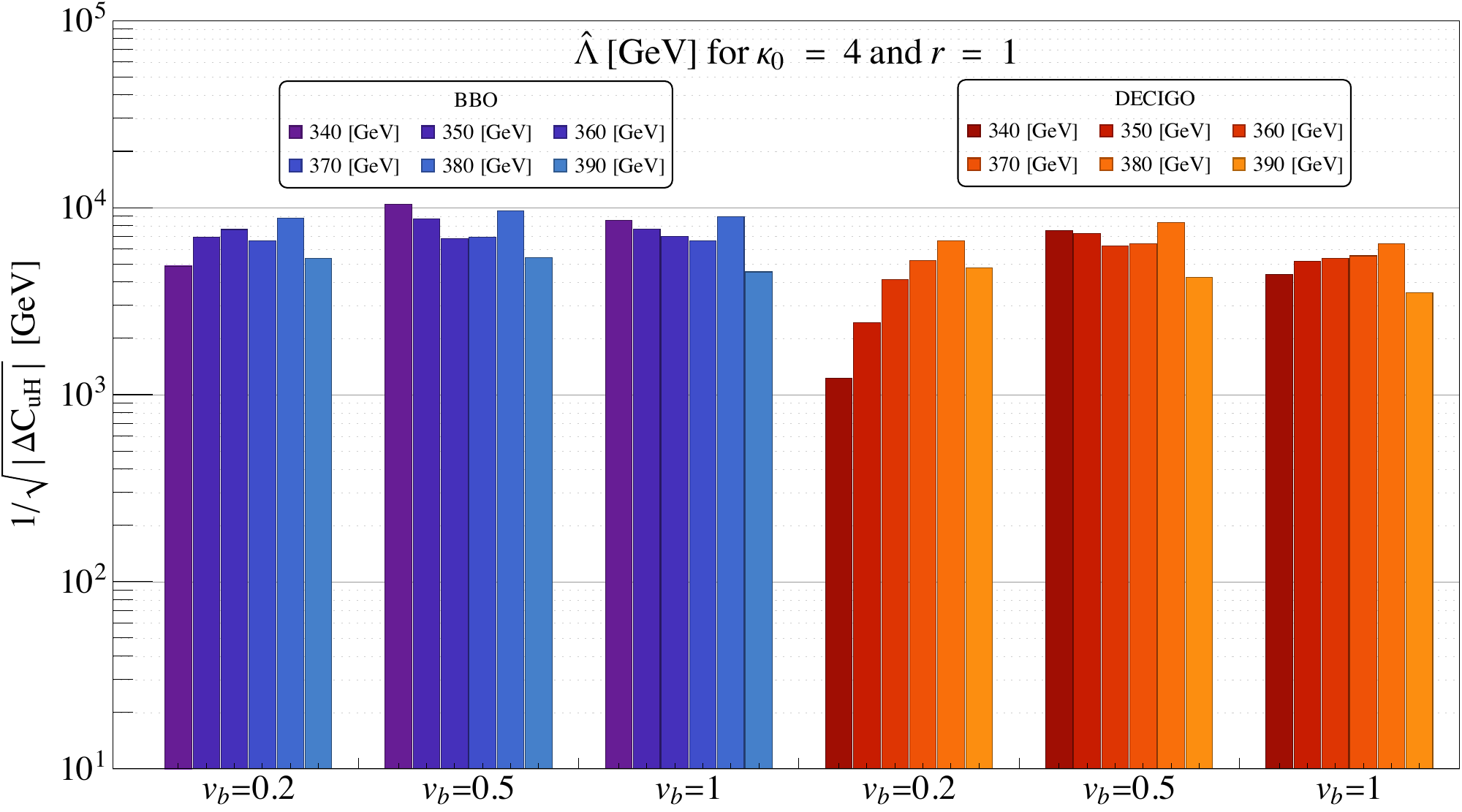}
\end{center}
\vspace{-0.4cm}
\caption{The same plots as Fig.~\ref{fig:k1r1} but for $(\kappa_0\,,\,r)=(4,\,1)$. 
}
\label{fig:k4r1}
\end{figure}

\begin{figure}[H]
\begin{center}
\includegraphics[width=16.cm]{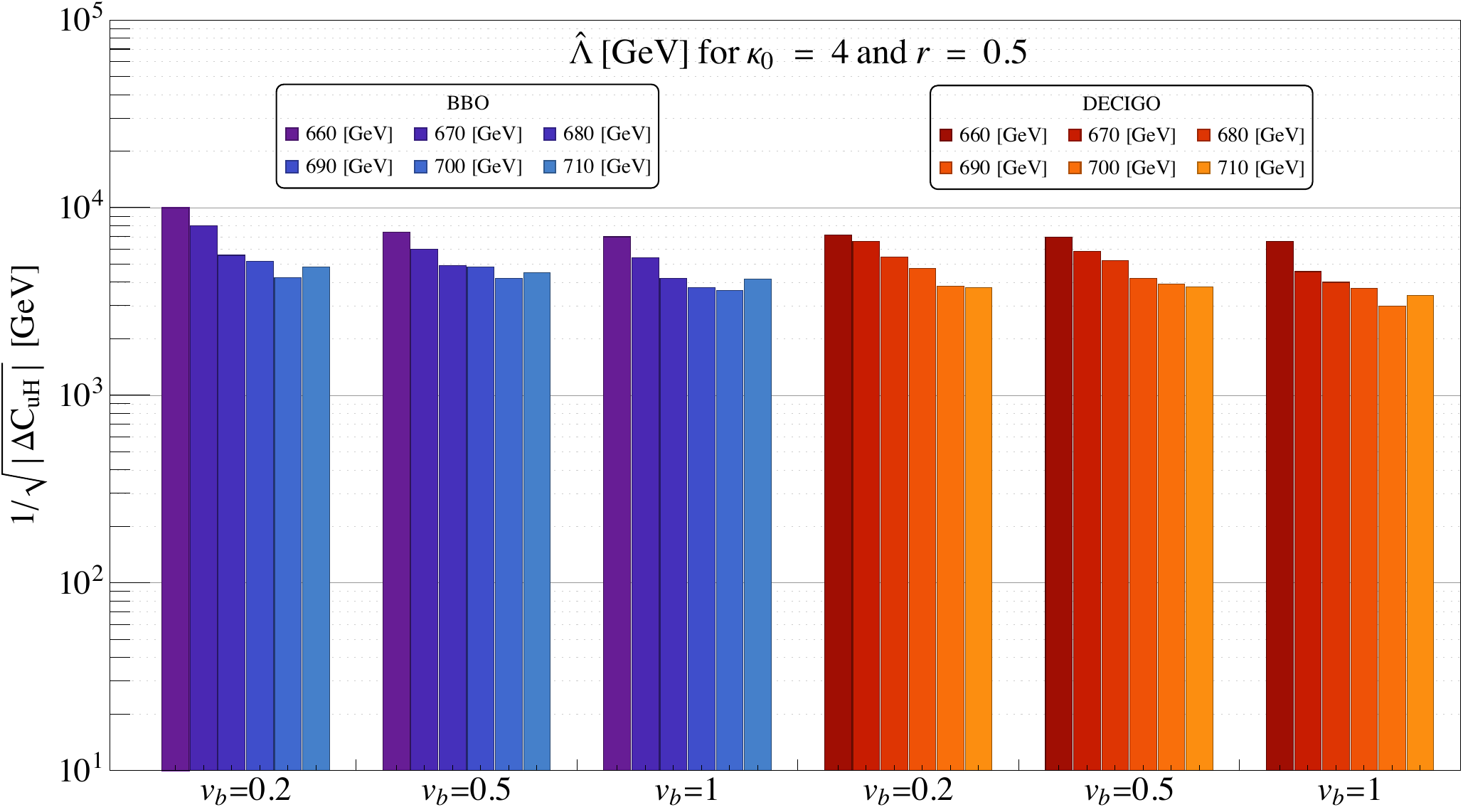}
\end{center}
\vspace{-0.4cm}
\caption{The same plots as Fig.~\ref{fig:k1r1} but for $(\kappa_0\,,\,r)=(4,\,0.5)$. 
}
\label{fig:k4r05}
\end{figure}

In Fig.~\ref{fig:k1r1}, we present the expected 95\% C.L. constraints for DECIGO (reddish bands) and BBO (bluish bands), assuming $\kappa_0=1$ and $r=1$ and considering the bubble-wall velocities $v_b=0.2$, $0.5$, and $1$.
The colored bands show the projected sensitivity to the effective scale $|\Delta C_{uH}|^{-1/2}$.
This sensitivity is obtained by taking the inverse square root of the upper bounds on $|\Delta C_{uH}|$ derived from the projections of the confidence ellipses onto the $\Delta C_{uH}$ axis, as illustrated in Fig.~\ref{fig:cont}.
Within each colored band, different shades correspond to different fiducial values of $\Lambda$, namely, $\hat{\Lambda}=500$, $510$, $520$, $530$, $540$, and $550~{\rm GeV}$.
These fiducial values are chosen from the regions of parameter space that yield detectable GW signals, as identified in Ref.~\cite{Hashino:2022tcs}.
The results show that, once a detectable GW signal is generated by non-decoupling NP, future GW observations can retain sensitivity to the benchmark coefficient $C_{uH}$.
Figures~\ref{fig:k1r05}, \ref{fig:k4r1}, and \ref{fig:k4r05} show analogous results for different ranges of $\hat{\Lambda}$, with the non-decoupling parameters $(\kappa_0,\,r)$ fixed to $(1,\,0.5)$, $(4,\,1)$, and $(4,\,0.5)$, respectively.
Taken together, Figs.~\ref{fig:k1r1}, \ref{fig:k1r05}, \ref{fig:k4r1}, and \ref{fig:k4r05} cover a broad range of the non-decoupling parameters $\Lambda$, $\kappa_0$, and $r$, focusing on the regions that yield detectable GW signals according to Ref.~\cite{Hashino:2022tcs}.
The projected effective NP scales are in the TeV range for many of the benchmarks and reach $\mathcal{O}(10)\,\mathrm{TeV}$ in favorable cases.
The reach can vary substantially with the benchmark parameters, including $\hat{\Lambda}$ and $v_b$, with some configurations yielding reaches below $1\,\mathrm{TeV}$. 
These results show that, within the benchmark scenarios and theoretical prescriptions adopted here, GW observations can be sensitive to the additional decoupling interaction associated with $C_{uH}$ even when the phase transition is driven by non-decoupling dynamics.
Therefore, the qualitative conclusions of Refs.~\cite{Hashino:2022ghd,Hashino:2025nku}, which assume that the GW signal is generated by the SMEFT dimension-six operator $(H^\dagger H)^3$ rather than by non-decoupling effects, remain applicable when the SFO-EWPT and the associated GW signal are instead driven by non-decoupling NP.
It should be emphasized, however, that these results demonstrate the potential sensitivity of GW observations rather than a complete determination of the underlying parameter space.
A robust interpretation of such constraints additionally requires knowledge of the fiducial values of $\Lambda$ and $(\kappa_0,\,r)$.
This would require improved control over the theoretical uncertainties in the prediction of the GW spectrum and, ultimately, complementary information on the parameters of the naHEFT framework from collider experiments (see, e.g., Ref.~\cite{Hashino:2025nku} for a detailed discussion).

\section{Summary and discussion}
\label{sec:summary}
In this work, we have investigated whether future gravitational-wave (GW) observations can probe additional decoupling new physics (NP) when a strong first-order electroweak phase transition (SFO-EWPT) is driven by non-decoupling dynamics. 
We described the non-decoupling effects using the nearly aligned Higgs effective field theory (naHEFT), while parametrizing the additional decoupling contribution by the dimension-six top–Higgs operator $(H^\dagger H)
(\overline{q}_3 t_R\widetilde{H})$.

Our numerical results show that, within the parameter regions considered, the strength of the phase transition is primarily controlled by the naHEFT contributions, while the additional contribution associated with \(C_{uH}\) produces comparatively small changes in \(v_c/T_c\).
Nevertheless, the resulting modifications to the phase-transition dynamics can leave potentially measurable imprints on the GW spectrum. 
Our Fisher matrix analysis indicates that DECIGO and BBO can retain sensitivity to $C_{uH}$.
The projected reach depends on the benchmark configuration, with effective NP scales reaching $\mathcal{O}(10)\,\mathrm{TeV}$ in favorable cases.
These results demonstrate that, within the naHEFT scenarios studied here, GW observations can retain sensitivity to additional decoupling effects even when the phase transition is driven by non-decoupling effects, rather than by the SMEFT operator $(H^\dagger H)^3$ as in Refs.~\cite{Hashino:2022ghd,Hashino:2025nku}.

For each benchmark, we performed a two-dimensional Fisher analysis in the \((\Lambda,C_{uH})\) plane, holding \(\kappa_0\), \(r\), and the bubble-wall velocity \(v_b\) fixed. We repeated this analysis for different choices of these parameters and of the fiducial value of \(\Lambda\). Although the projected reach varies with the benchmark, the qualitative conclusion that GW observations retain sensitivity to the additional decoupling contribution holds across the benchmark choices considered. The quoted sensitivities describe the expected precision of the two-parameter fit at each benchmark. 
Appendix~\ref{sec:rge} illustrates the renormalization-scale dependence of the projected sensitivity for a representative benchmark.
A more systematic assessment of theoretical uncertainties in the effective potential and GW spectrum, including the dependence on the thermal resummation prescription, would help establish the robustness of the projected sensitivities.

The effects of other decoupling operators can be evaluated straightforwardly by adapting the results of our previous studies~\cite{Hashino:2022ghd,Hashino:2025nku} to the present framework. Given the large number of such operators, we have focused on $(H^\dagger H)
(\overline{q}_3 t_R\widetilde{H})$ as a representative benchmark. Applying the same framework to other operators would allow us to compare the sensitivity of GW observations to different interactions. In addition, collider measurements of the Higgs self-coupling and top Yukawa coupling could provide complementary constraints on the parameters entering the GW predictions. Combining these constraints with GW observations could help distinguish parameter choices that produce similar GW spectra.

\section*{Acknowledgements}
DU is supported by grants from the ISF (No.~1002/23 and 597/24) and the BSF (No.~2021800).
This work is supported by Grant-in-Aid for Early-Career Scientists 25K17398 (KH). 

\appendix

\section{RGE effects}
\label{sec:rge}

\begin{figure}[H]
\begin{center}
\includegraphics[width=10.cm]{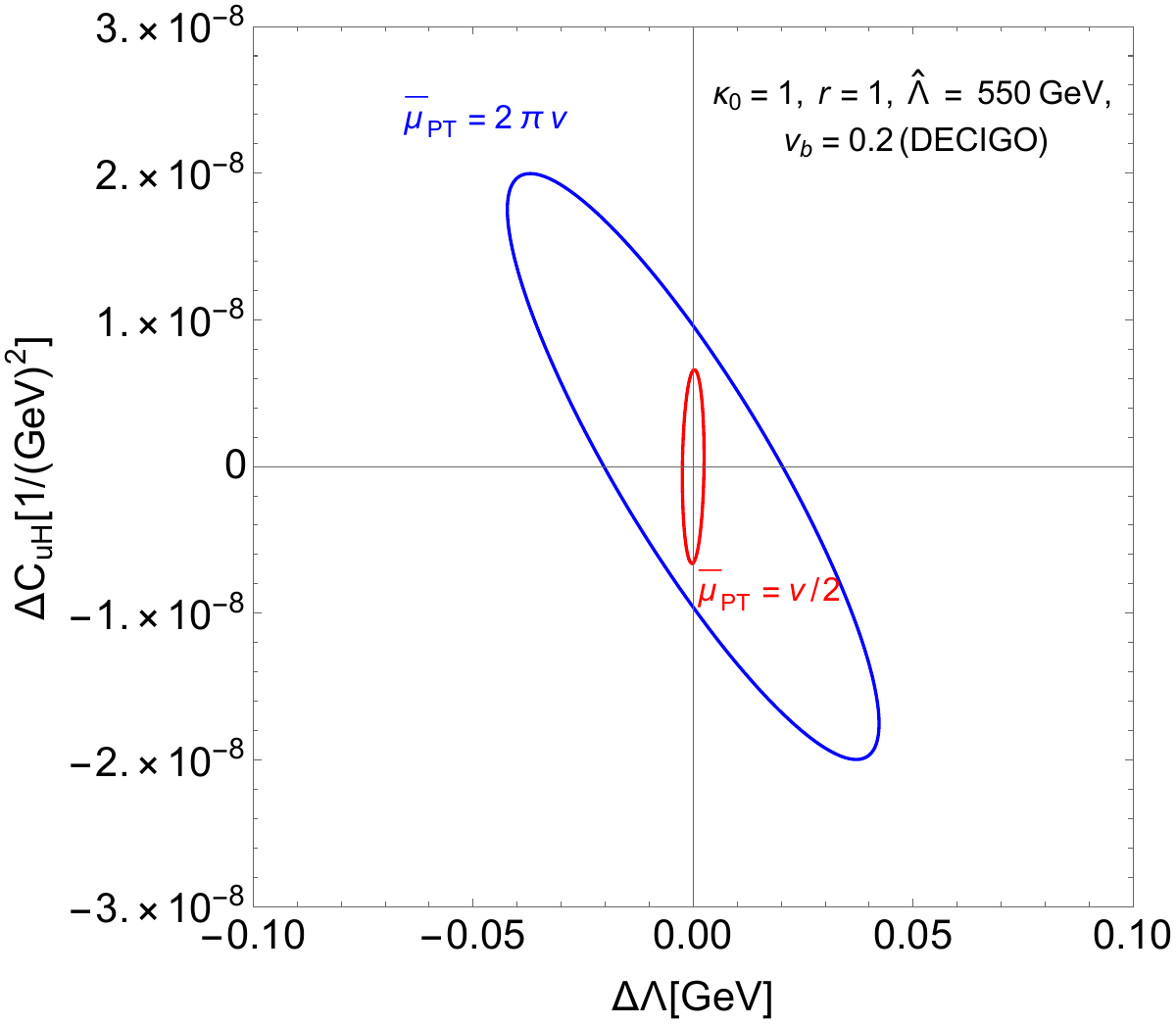}
\end{center}
\vspace{-0.4cm}
\caption{
Expected 95\% joint confidence contours for DECIGO, assuming an observation time of one year.
The blue and red contours correspond to the renormalization scales $\bar{\mu}_{\rm PT}=2\pi v$ and $\bar{\mu}_{\rm PT}=v/2$, respectively.
In both cases, the fiducial point is $(\hat{\Lambda},\,\hat{C}_{uH})=(550~{\rm GeV},\,0)$, with $(\kappa_0,\,r)=(1,\,1)$ and $v_b=0.2$ held fixed.
}
\label{fig:rge}
\end{figure}

In this appendix, we examine the impact of renormalization-scale dependence on our Fisher matrix forecasts for a representative benchmark, following Ref.~\cite{Hashino:2025nku}.
Even in the present framework, in which non-decoupling effects drive the SFO-EWPT, this dependence can affect the projected sensitivity to $C_{uH}$ and its correlation with $\Lambda$.
To illustrate this dependence, Fig.~\ref{fig:rge} compares the expected 95\% joint confidence regions for DECIGO in the $(\Delta\Lambda,\,\Delta C_{uH})$ plane, assuming an observation time of one year.
The blue and red contours correspond to the renormalization scales $\bar{\mu}_{\rm PT}=2\pi v$ and $\bar{\mu}_{\rm PT}=v/2$, respectively.
See Ref.~\cite{Hashino:2025nku} for details of the procedure used to assess the dependence on the renormalization scale $\bar{\mu}_{\rm PT}$.
Although we employ the Parwani scheme rather than the thermal resummation prescription adopted in Ref.~\cite{Hashino:2025nku}, we follow the same procedure for the renormalization-group evolution of the parameters.
In both cases, we fix $(\kappa_0,r)=(1,1)$ and adopt the fiducial point $(\hat{\Lambda},\,\hat{C}_{uH})=(550~{\rm GeV},\,0)$.
This result suggests that the theoretical uncertainties arising from renormalization-scale dependence in the present Fisher matrix analysis are qualitatively similar to those discussed in Ref.~\cite{Hashino:2025nku}.
A more comprehensive assessment of renormalization-scale uncertainties in the present framework is left for future work.

\section{Field-dependent masses}
\label{sec:FDM}
We present below the field-dependent masses of the Higgs boson, the NG modes, the gauge bosons, and the top quark.
To simplify the expressions, we first define
\begin{align}
    A_\phi
    &\equiv
    -m^2+\frac{\lambda}{2}\phi^2
    .
    \label{eq:FDM_aux}
\end{align}
The field-dependent squared masses are then given by
\begin{align}
    M_h^2(\phi)
    &=
    A_\phi+\lambda\phi^2\,,
    \label{eq:FDM_h}
    \\
    \bigl(M_1^\pm(\phi)\bigr)^2
    &=
    \frac{1}{2}\left[
        A_\phi
        \pm
        \sqrt{A_\phi}\,
        \sqrt{
            A_\phi
            -\left(g'^2\xi_B+g^2\xi_W\right)\phi^2
        }
    \right]\,,
    \label{eq:FDM_1}
    \\
    \bigl(M_2^\pm(\phi)\bigr)^2
    &=
    \frac{1}{2}\left[
        A_\phi
        \pm
        \sqrt{A_\phi}\,
        \sqrt{A_\phi-g^2\xi_W\phi^2}
    \right]\,,
    \label{eq:FDM_2}
    \\
    M_W^2(\phi)&=\frac{1}{4}\,g^2 \phi^2\,,\label{eq:MW_phi}
    \\
    M_{Z}^2(\phi)&=\frac{1}{4}\left({{g'}}^2+{g}^2\right)\phi^2\,,
    \\
    M_\gamma^2(\phi)&=0\,,\label{eq:M2_phi}
    \\
    M_t^2(\phi)&=\frac{1}{2}\phi^2 \left(Y_t -\frac{1}{2}\phi^2 \,C_{u H}\right)^2\,.\label{eq:Mt_phi}
\end{align}

\section{Thermally resummed masses}
\label{sec:resum}
The thermally resummed masses appearing in Eq.~\eqref{eq:Vfull} are given by
\begin{align}
    &M^2_{W,{\rm res.}}(\phi,T)=M^2_W(\phi)+m_{\rm D}^2\,,
    \\
    &M^2_{Z,{\rm res.}}(\phi,T)= \frac{1}{8}\Bigg[
    ({g'}^2+{g}^2)\phi^2
    +4 (m_{\rm D}^2+{m'_{\rm D}}^2)\notag
    \\
    &\qquad+4\sqrt{\left(m_{\rm D}^2+{m'}^2_{\rm D}+\frac{({g'}^2+{g}^2)\phi^2}{4}\right)^2-4 m_{\rm D}^2 {m'}^2_{\rm D}- (m_{\rm D}^2 {g'}^2+{m'}^2_{\rm D}{g}^2 )\phi^2}
    \Bigg]\,,
    \\
    &M^2_{A',{\rm res.}}(\phi,T)=
    \frac{1}{8}\Bigg[
    ({g'}^2+{g}^2)\phi^2
    +4 (m_{\rm D}^2+{m'_{\rm D}}^2)\notag
    \\
    &\qquad-4\sqrt{\left(m_{\rm D}^2+{m'}^2_{\rm D}+\frac{({g'}^2+{g}^2)\phi^2}{4}\right)^2-4 m_{\rm D}^2 {m'}^2_{\rm D}- (m_{\rm D}^2 {g'}^2+{m'}^2_{\rm D}{g}^2 )\phi^2}
    \Bigg]
    \,,
    \\
    &M^2_{h,{\rm res.}}(\phi,T)=\mu^2_{{\rm res.}}+3 \lambda_{\rm res.}\phi^2\,
    ,
    \\
    &M^2_{1\pm,{\rm res.}}(\phi,T)=\frac{1}{2}\Bigg[m_{\chi,{\rm res.}}^2\pm\sqrt{m_{\chi,{\rm res.}}^2\left(m_{\chi,{\rm res.}}^2-({g'}^2\xi_B+{g}^2\xi_W)\phi^2\right)}\Bigg]\,,
    \\
    &M^2_{2\pm,{\rm res.}}(\phi,T)=\frac{1}{2}\Bigg[m_{\chi,{\rm res.}}^2\pm \sqrt{m_{\chi,{\rm res.}}^2 \left(m_{\chi,{\rm res.}}^2-\xi_W {g}^2\phi^2 \right)}\Bigg]\,,
    \\
    &m_{\chi,{\rm res.}}^2=\mu^2_{{\rm res.}}+\lambda_{\rm res.} \phi^2\,,
    \\
    &m_{\rm D}^2 =T^2 g^2 \left(\frac{5}{6}+\frac{1}{3}N_f\right)\,,
    \\
    &{m'}_{\rm D} ^2=T^2 {g'}^2 \left(
    \frac{1}{6} +\frac{5}{9}N_f
    \right)\,,
    \\
    &\mu^2_{{\rm res.}}=-m^2 +\frac{T^2}{12}\left(
    3 \lambda +\frac{3}{4} \left(3 g^2 +{g'}^2\right)
    +3 Y_t^2
    \right)\,,
    \\
    &\lambda_{\rm res.}=\frac{\lambda}{2}\,,
\end{align}
where $N_f=3$ denotes the number of kinematically active families at electroweak-scale temperatures.
See also Ref.~\cite{Croon:2020cgk}.
In evaluating the Higgs and NG contributions to the thermal potential, we omit the explicit naHEFT-induced corrections to their field-dependent masses and thermal self-energies.
The explicit naHEFT-induced corrections to the Higgs and NG field-dependent masses are also omitted in the CW potential, since their insertion would generate contributions of formal two-loop order or higher.

\clearpage
\bibliography{PT2.bib}

\end{document}